\documentclass[aip,amsmath,amssymb,reprint,]{revtex4-1}

\usepackage{graphicx}
\usepackage{dcolumn}
\usepackage{bm,upgreek}

\usepackage[utf8]{inputenc}
\usepackage[T1]{fontenc}
\usepackage{mathptmx}

\begin{document}

\preprint{AIP/123-QED}

\title{Strong magnon--photon coupling within a tunable cryogenic microwave cavity}

\author{C.A. Potts}
\affiliation{Department of Physics, University of Alberta, Edmonton, Alberta T6G 2E9, Canada}
\email{cpotts@ualberta.ca, jdavis@ualberta.ca}

\author{J.P. Davis}
\affiliation{Department of Physics, University of Alberta, Edmonton, Alberta T6G 2E9, Canada}%

\date{\today}

\begin{abstract}

The ability to achieve strong-coupling has made cavity-magnon systems an exciting platform for the development of hybrid quantum systems and the investigation of fundamental problems in physics. Unfortunately, current experimental realizations are constrained to operate at a single frequency, defined by the geometry of the microwave cavity. In this article we realize a highly-tunable, cryogenic, microwave cavity strongly coupled to magnetic spins. The cavity can be tuned \textit{in situ} by up to 1.5 GHz, approximately 15\% of its original 10 GHz resonance frequency. Moreover, this system remains within the strong-coupling regime at all frequencies with a cooperativity of $\approx 800$.  

\end{abstract}

\maketitle

Since the first demonstrations of strong magnon--photon coupling,\cite{Tabuchi_2014,Zhang_2014} the field of cavity-magnonics has become increasingly diverse and fruitful. This is largely due to the relative ease of achieving strong-coupling (\textit{i.e.}, the coupling rate is larger than the dissipation rate of either sub-system), resulting from the large spin density of yttrium iron garnet (YIG). YIG has been the material of choice for many magnonic experiments because of its high spin density, but also because it exhibits low spin damping and good optical properties.\cite{Stancil_Spin_2000} The commercial availability of high-quality YIG samples,\cite{FerriSphere}  has allowed the establishment and rapid growth of the field of cavity-magnonics.\cite{Morris_2017,Bai_Spin_2015,viennot_coherent_coupling_2015,Cao_Exchange_2015,zhang_magnon_dark_2015,Gloppe_Resonant_2019,Goryachev_2018,Bhoi_2014,Li_2019} Some notable recent experiments include the study of dissipative coupling, namely the observation of exceptional points characterized by level-attraction,\cite{Harder_2018} as well as nonreciprocity and unidirectional invisibility.\cite{Wang_2019} The magnetostrictive interaction has also been observed, coupling magnons to the breathing phonon modes of spherical YIG samples,\cite{Zhang_2016} which enabled our theoretical proposal for a primary thermometer based on a cavity-magnonic system.\cite{Potts_2020} Extensive progress has been made in the use of cavity-magnoics for the development of hybrid quantum technologies. These include the coupling of magnons to superconducting qubits,\cite{Tabuchi_2015}  bi-directional conversion between microwave and optical photons,\cite{Hisatomi_2016,Zhu_2020} and heralded generation of single magnons.\cite{Lachance-Quirion_2020} Finally, there has been a recent proposal for direct detection of axions---a dark matter candidate particle---by probing, with a microwave cavity, the electron-axion coupling, effectively described as an oscillating magnetic field acting on the magnons of the material.\cite{Flower_2019,Crescini_2020}

In the majority of these experiments, the magnonic mode is coupled to the magnetic field of a microwave cavity and, as the resonance frequencies of these magnonic modes within YIG depend linearly on the applied static magnetic field, provides a large degree of tunability. Unfortunately, with current experimental architectures, the microwave resonance frequency is not tunable, limiting the possibility of bringing a coupled cavity-magnonic system into resonance with additional sub-systems. The inability to independently tune both the magnon and microwave cavity resonances has detrimental effects on total system efficiency, for example, when coupling magnons to a superconducting qubit or in a microwave-to-optical transduction experiment. Furthermore, there may exist experimental setups in which a tunable magnetic field is not practical to implement. Therefore, it would be advantageous to have a microwave cavity with the ability to \textit{in situ} tune its resonance frequency,\cite{Hyde_2017,Bourhill_2019_spectroscopy} especially while at cryogenic temperatures. 

In this letter, we solve this problem by demonstrating a highly-tunable, cryogenic, microwave cavity strongly coupled to the lowest-order ferromagnetic resonance within a YIG sphere. Our cavity design is based upon a double-stub re-entrant cavity, similar to those described in Refs.~\citenum{Flower_2019,Goryachev_2018}. This tunable hybrid system reaches the strong coupling regime ($g/\{\kappa,\gamma\} > 1$), and is nearly within the ultra-strong coupling (USC) regime ($g/\omega>0.1$). This architecture allows the implementation of hybrid quantum technology with three distinct modes (microwave, magnonic, and auxiliary) that can be simultaneously brought into resonance.

\begin{figure}[b]
\includegraphics[width= 7.5 cm]{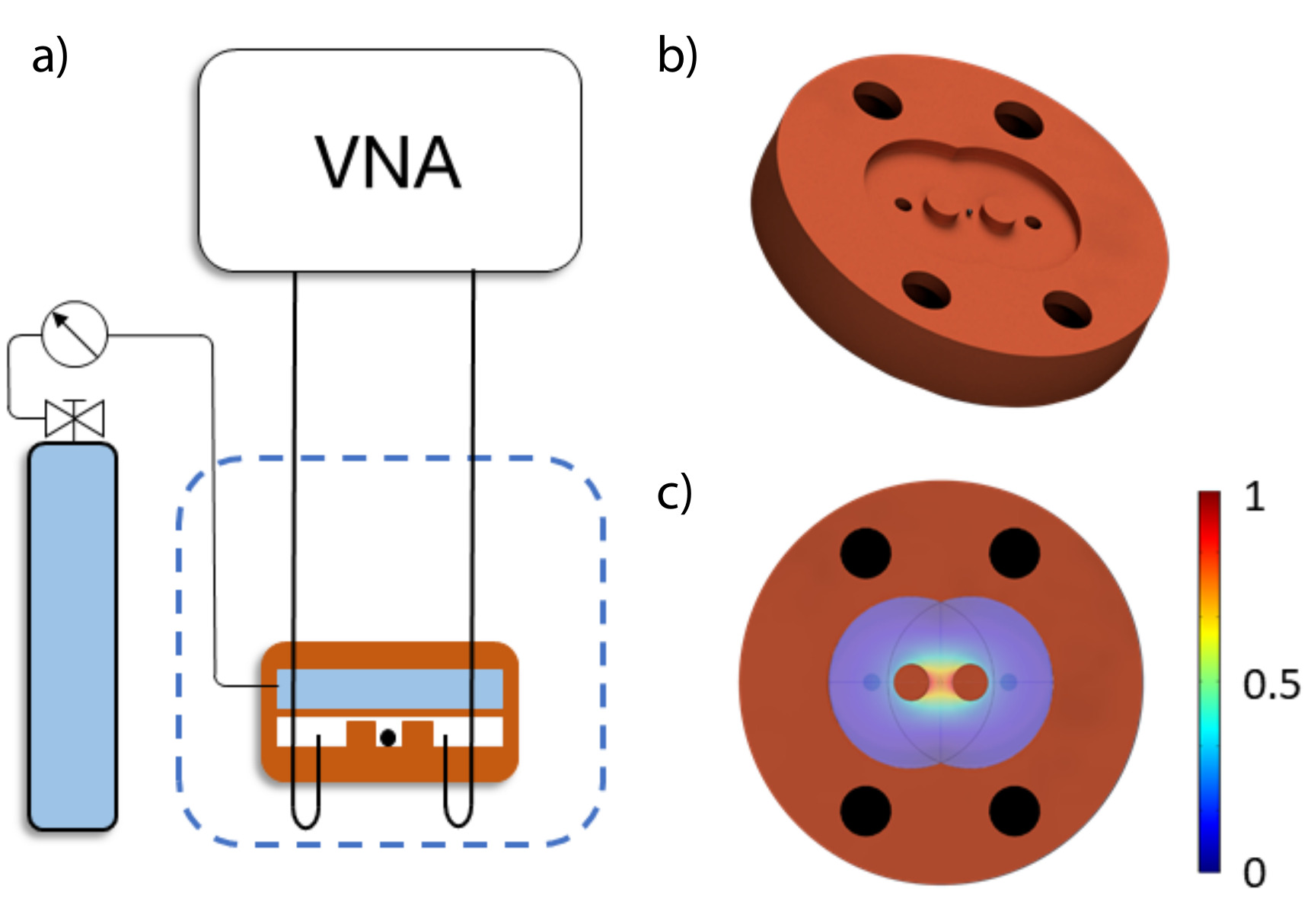}
\caption{a) A schematic outlining the experimental apparatus. A helium reservoir is connected to the cavity through a pressure control valve. Transmission scans are performed using a VNA coupled to the cavity via pin couplers inserted into the cavity. b) 3D rendering of the microwave resonator. The $500$ $\mathrm{\upmu m}$ diameter YIG sphere lies between the two re-entrant posts in the high-magnetic-field region.  c) A finite element simulation of the anti-symmetric magnetic field mode profile, localized between the re-entrant posts. }
\label{Fig:01}
\end{figure}

The cavity (as seen in Fig.~\ref{Fig:01}) consists of two partially overlapping 14 mm diameter circles, milled 1 mm deep. The re-entrant stubs are 3 mm in diameter and located on the axis of the two cylindrical depressions separated by a 2 mm gap. A commercially available YIG sphere, $500$ $\mathrm{\upmu m}$ in diameter, is placed directly between the two posts. This location provides the maximum overlap while maintaining a relatively uniform magnetic field over the entire volume of the YIG sphere. Double-stub re-entrant cavities of this form support two resonant modes, a lower frequency symmetric mode, and a higher frequency antisymmetric mode. The magnetic field of the symmetric mode is expelled from between the posts and is ignored here. Instead, we focus on the higher-frequency antisymmetric mode; this mode focuses the magnetic field between the posts as shown in Fig.~\ref{Fig:01}c. By placing the small YIG sphere between the posts, the mode overlap between the YIG sphere and the magnetic field can be large. This large mode overlap enhances the coupling between the cavity field and the magnetic sample.

Both halves of the double-stub reentrant cavity were machined out of oxygen-free high-conductivity copper. A model of the bottom half of the cavity is shown in Fig.~\ref{Fig:01}b. The two circular posts have been machined to leave a small gap ($d\approx 100$ $\mathrm{\upmu m}$) between them and the flat membrane comprising the other half of the cavity. The thin, $\sim 500$ $\mathrm{\upmu m}$, membrane on the top half of the cavity is backed by a reservoir of liquid helium, fed by a thermally anchored capillary. The pressure within the helium reservoir is controlled via a pressure regulator at room temperature (Fig.~\ref{Fig:01}a).\cite{Clark_2018} Increasing the pressure causes the membrane to deform, reducing the distance between the membrane and the two circular posts. The tunability of this cavity can be understood by noticing that the electric field is primarily confined within this gap. This small gap can be approximated as a parallel plate capacitor, forming the capacitance of a lumped model LC circuit representing the microwave cavity. The effective capacitance can be approximated as $C = \epsilon_0 A/d$, where $\epsilon_0$ is the permittivity of free space, $A$ is the total area of the posts, and $d$ is the distance between the capacitor plates. Using this assumption, and assuming the inductance $L$ of the circuit is not affected by the deformation of the membrane, it is straightforward to see that the resonance frequency, $\omega = 1/\sqrt{LC}$, is proportional to the square root of the distance between the membrane and the posts. From this simple model, the cavity’s tunability is understood as follows; as the pressure inside the reservoir is increased, the gap between the membrane and the posts is reduced and the resonance frequency is decreased. 

\begin{figure}[b]
\includegraphics[width=7.5 cm]{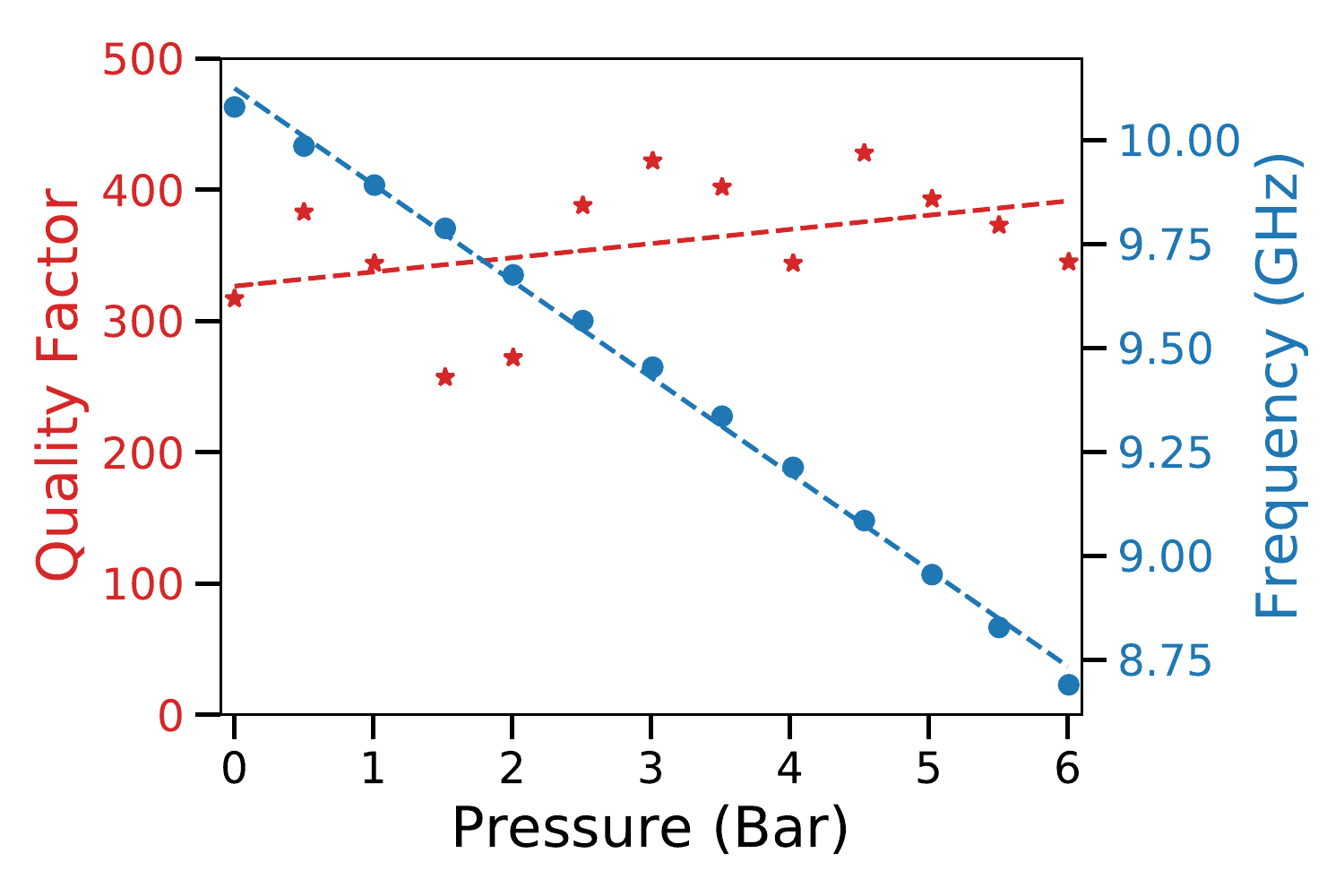}
\caption{Quality factor and center frequency of the microwave cavity loaded with a YIG sphere. Linear fit reveals a slow increase in the quality factor as the reservoir pressure is increased. Data was taken with zero applied static magnetic field. }
\label{Fig:02}
\end{figure}

The copper microwave cavity was attached to the base plate of a cryostat operating at 1.2 K for the duration of the measurements. The microwave cavity mode was driven via a coaxial cable, thermally anchored at 4 K. The readout was performed in transmission ($S_{21}(\omega)$) of a vector network analyzer (VNA), Fig.~\ref{Fig:01}a) with no amplification, and as a result the measurements were performed with relatively high input power (approximately -20 dBm at the input port). External coupling to the cavity was achieved using straight pin couplers, however, due to the large electric field confinement, was highly undercoupled. If critical coupling was required one could use loop-couplers, as demonstrated by Clark \textit{et al.}~in Ref.~\citenum{Clark_2018}.  Characterization of the cavity tunability was performed by measuring the resonance frequency and quality factor as a function of reservoir pressure, shown in Fig.~\ref{Fig:02}. The primary factor limiting the quality factor of this cavity is the seam that exists between the two halves of the cavity. The quality factor may be increased in future implementations by  bonding the two halves of the cavity using a galvanic indium bond.\cite{Brecht_2015}

As described by Walker,\cite{ Walker_1958} and further elaborated by Fletcher \textit{et al.},\cite{Fletcher_1959} ferromagnetic spheres contain many magnetostatic resonances. For this work, we are strictly interested in coupling to the uniformly processing Kittel mode. The frequency of the Kittel mode is determined by the magnitude of an applied static magnetic field, applied here by a superconducting magnet surrounding the sample.  Specifically, it depends linearly on the magnitude of the applied field, $\omega_m = \gamma \vert B_0 \vert + \omega_{m,0}$, where $\gamma = 28 $ GHz/T is the gyromagnetic ratio and $\omega_{m,0}$ is an offset determined by the anisotropy field. Moreover, YIG has a cubic magneto-crystalline anisotropy, with an easy magnetization axis along the (111) direction. However, for this experiment we were not concerned with any intrinsic anisotropy effect and have randomly oriented the YIG sample within the cavity. This random orientation nevertheless results in a shift in the ferromagnetic resonances due to induced anisotropy fields. 

\begin{figure}[b]
\includegraphics[width= 7.5 cm]{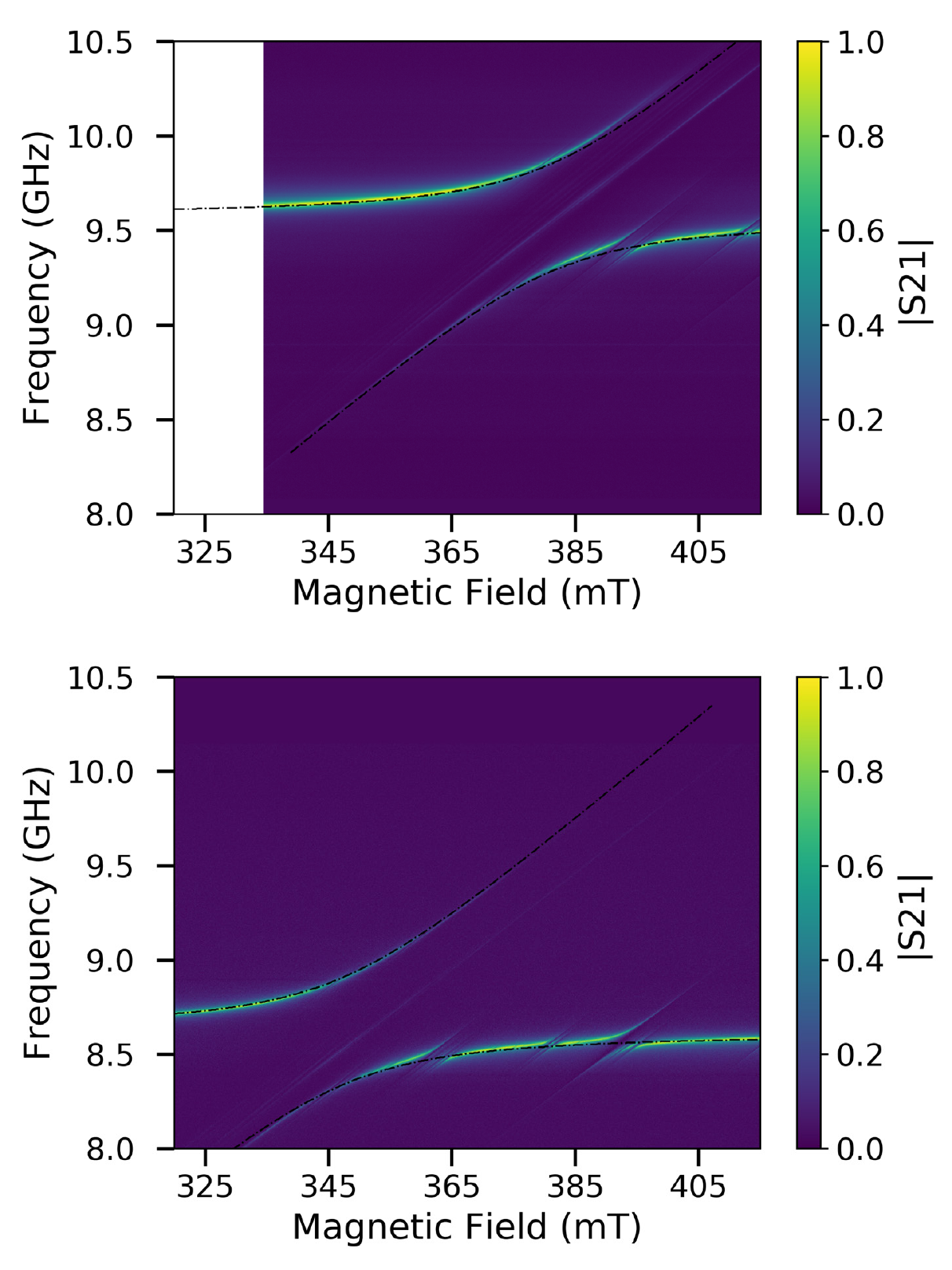}
\caption{a) Normalized transmission spectrum ($S_{21}(\omega)$) of the tunable YIG microwave cavity system as a function of external magnetic field. The helium reservoir pressure was set to 2 bar. Dotted black line is a fit to the coupled magnon-photon system described by Eq.~\eqref{eqn_s21}. b) Normalized transmission spectrum ($S_{21}(\omega)$), for a helium reservoir pressure set to 6 bar. Note, both figures have been plotted on the same axes.}
\label{Fig:03}
\end{figure}

First, we characterize the magnon-photon interaction at various cavity frequencies. The cavity frequency is set by pressurizing the helium reservoir, and is held constant over the course of an experimental run. The magnetic field was slowly increased such that the magnon frequency passes through the cavity frequency, thus bringing the magnons and photons into resonance. At each static magnetic field step, transmission measurements were performed. The resulting scattering parameter ($S_{21}(\omega)$) is plotted in Fig.~\ref{Fig:03}a,b for two pressures, 2 and 6 bar, respectively. One can see the avoided level crossing, which is the hallmark of mode hybridization and strong-coupling.  

The interaction between the magnons and photons can be described by the following Hamiltonian,\cite{Tabuchi_2014,Zhang_2014}
\begin{equation}
     \hat{\mathcal{H}} = \hbar \omega_{\textrm{c}} \hat{a}^{\dagger}\hat{a} + \hbar \omega_{\textrm{m}} \hat{m}^{\dagger}\hat{m} + \hbar g (\hat{a}^{\dagger}\hat{m} + \hat{a}\hat{m}^{\dagger}).
\end{equation}
\noindent Here $\hat{a}^{\dagger}$ ($\hat{a}$) is the creation (annihilation) operator for the microwave photons at frequency $\omega_c$. The magnons---with resonance frequency $\omega_m$---are represented by the bosonic operators $\hat{m}^{\dagger}$ ($\hat{m}$) via the Holstein-Primakoff approximation.\cite{Holstein_1940} The photon-magnon coupling rate is given by,
\begin{equation}
    g = \frac{\gamma}{2}\sqrt{\frac{\mu_{\textrm{0}} \hbar \omega}{V_{\textrm{eff}}}}\sqrt{2Ns}.
    \label{eqn_g}
\end{equation}
\noindent Here, $\omega$ is the resonance frequency, $\mu_0$ is the vacuum permeability, $N=  4/3 \rho \pi r^3$ is the total number of spins (where $r$ is the radius of the YIG sphere), $\rho = 4.22 \times 10^{27}$ m$^{-3}$ is the spin density of YIG, and $s=5/2$ is the spin number for the Fe$^{3+}$ ions within YIG. Further, $V_\textrm{eff}$ is the effective mode volume of the microwave cavity, defined as,
\begin{equation}
    V_{\textrm{eff}} = \int \frac{\mu(\textbf{r})\vert \textbf{B}(\textbf{r}) \vert^2}{\mu(\textbf{r}_{\textrm{0}})\vert \textbf{B}(\textbf{r}_{\textrm{0}}) \vert^2} dV.
\end{equation}
\noindent This integral is performed over the volume of the microwave cavity and $\textbf{r}_0$ is the location of the YIG sphere within the cavity. It should be noted that Eq.~\eqref{eqn_g} is valid only when the microwave magnetic field is approximately constant over the entire magnetic sample. If the magnetic field has large variation over the sample volume one may use the equations for the coupling rate defined in Ref.~\citenum{Flower_Experimental_2019}.

The magnon--photon coupling rate and the magnon linewidth can be extracted from the experimentally measured transmission coefficient, $S_{21}(\omega)$, where the transmission coefficient can be evaluated using input-output theory,\cite{gardiner_quantum_2000} and is given explicitly by\cite{Tabuchi_2014}
\begin{equation}
    S_{21}(\omega) = \frac{\sqrt{\kappa_1\kappa_2}}{i(\omega - \omega_c)- \frac{\kappa}{2}+ \frac{\vert g \vert}{i(\omega-\omega_{\textrm{m}})-\gamma_{\textrm{m}}/2}}.
    \label{eqn_s21}
\end{equation}
\noindent Here, $\kappa_i$ is the coupling rate to the $i$-th external port, and $\kappa = \kappa_1+\kappa_2+\kappa_{\textrm{int}} \sim \kappa_{\textrm{int}}$ is the total cavity decay rate, which is approximately equal to the internal loss rate since the coupling ports are highly undercoupled. The magnon linewidth is $\gamma_m$. The peak of the theoretical curve is plotted on top of the experimental data in Fig.~\ref{Fig:03}a,b. Using Eq.~\eqref{eqn_s21} we can extract the magnon-photon coupling rate $g/2\pi = 285$ MHz, which is in good agreement with the value extracted from COMSOL via Eq.~\eqref{eqn_g}, $g/2\pi = 260$ MHz. Furthermore, we can extract the magnon linewidth $\gamma_m = 4.3$ MHz, and the total cavity linewidth $\kappa/2\pi = 23.8$ MHz.

\begin{figure}[t]
\includegraphics[width= 7.5 cm]{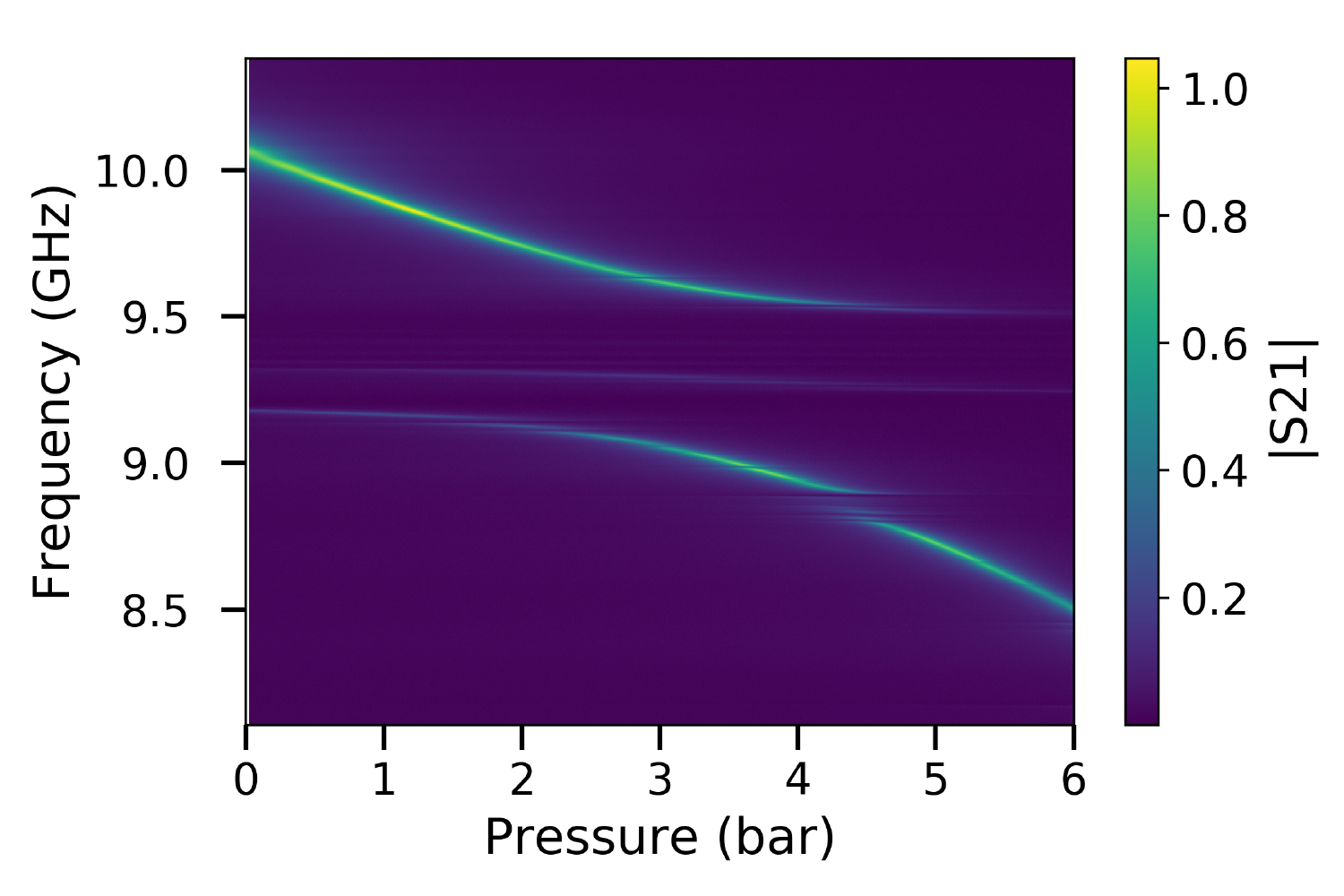}
\caption{ Normalized transmission spectrum ($S_{21}(\omega)$) of the tunable YIG microwave cavity system as a function of helium reservoir pressure. The magnetic field was held constant at $\sim 380$ mT. }
\label{Fig:04}
\end{figure}

From these parameters it can be seen that this cavity lies well within the strong-coupling regime ($g/\{\kappa,\gamma_{\textrm{m}}\} >1$) and is close to the USC regime ($g/\omega > 0.1$). It would be possible to reach the USC regime by using a larger YIG sphere. Based on COMSOL simulations a $1$ mm diameter YIG sphere should be well within the USC regime. Furthermore, as can be seen in Fig.~\ref{Fig:03} and Fig.~\ref{Fig:04} there exists faint transmission between the hybridized cavity-magnon modes. This behavior was predicted by Rameshti \textit{et al.}, and may be understood as nearly unperturbed higher-order magnon modes that couple to the cavity as a result of ultra-strong coupling effects.\cite{Rameshti_2015} Similar observation have been made in previous experiments within the USC regime.\cite{Bourhill_2016,Zhang_2014}

From the experimentally extracted parameters we can determine the cooperativity for this experimental setup to be $C = g^2/\kappa\gamma_{\textrm{m}} \approx 800$. Since the cooperativity scales with the radius cubed, to compare our results with the literature we shall considered the cooperativity per volume. In our experiment, we have achieved $C_{\textrm{V}_{\textrm{m}}} = 1210$ mm$^{-3}$, a value comparable to the volume-normalized cooperativity obtained in Ref.~\citenum{Bourhill_2016} and an order of magnitude below the state-of-the-art $C_{\textrm{V}_{\textrm{m}}} = 56000$ mm$^{-3}$.\cite{Goryachev_2014} The difference between our results and state-of-the-art values can be minimized with optimization of our microwave cavity dissipation, as described above.

A final demonstration of the tunability of our microwave cavity was performed by examining the magnon-photon strong-coupling while \textit{not} varying the magnitude of the static magnetic field. This experimental procedure is different than typical cavity-magnonic experiments in which the cavity frequency is fixed and the magnon frequency is swept, as is shown in Fig.~\ref{Fig:03}. Instead, we set the static magnetic field to a constant value of $\sim 380$ mT. Then the pressure of the helium reservoir was slowly reduced from 6.0 bar to 0 bar over the course of several hours. During the pressure sweep transmission scans were constantly performed. The resulting transmission data is shown in Fig.~\ref{Fig:04}. This plot is similar to what is seen in Fig.~\ref{Fig:03}, however, now the cavity frequency shifts, while the magnon modes are held at a constant frequency. This additional degree of freedom will allow the frequency of maximum hybridization ($\omega_c = \omega_m$) to be set anywhere within the range of the tunability of the microwave cavity, rather than being restricted to a single frequency set by the microwave cavity. This feature is especially important at cryogenic temperatures where the cavity frequency may shift during cooling as a result of thermal contraction.

In conclusion, we have demonstrated a tunable cryogenically-compatible 3D microwave cavity strongly coupled to magnetic spin excitation of a YIG sphere. The cavity can be tuned over a range of 1.5 GHz while maintaining an approximately constant quality factor. Furthermore, we demonstrate a mechanism for studying magnons that provides complete experimental control over the frequency of maximal magnon-photon hybridization \textit{in situ} at cryogenic temperatures. This cavity design could provide improvements for a variety of cavity magnonic experiments ranging from quantum information experiments such as microwave-to-optical transaction, and---with optimization---to fundamental problems in physics, such as the search for dark matter candidates.

\begin{acknowledgments}
Authors acknowledge helpful contributions from V. Vadakkumbatt and E. Varga. This work was supported by the University of Alberta; the Natural Sciences and Engineering Research Council, Canada (Grants No. RGPIN-04523-16, No. DAS-492947-16, and No. CREATE-495446-17); and Alberta Quantum Major Innovation Fund.  Data are available from the corresponding author upon reasonable request.
\end{acknowledgments}

\nocite{*}
\bibliography{aipsamp}

\providecommand{\noopsort}[1]{}\providecommand{\singleletter}[1]{#1}%
\begin{thebibliography}{35}%
\makeatletter
\providecommand \@ifxundefined [1]{%
 \@ifx{#1\undefined}
}%
\providecommand \@ifnum [1]{%
 \ifnum #1\expandafter \@firstoftwo
 \else \expandafter \@secondoftwo
 \fi
}%
\providecommand \@ifx [1]{%
 \ifx #1\expandafter \@firstoftwo
 \else \expandafter \@secondoftwo
 \fi
}%
\providecommand \natexlab [1]{#1}%
\providecommand \enquote  [1]{``#1''}%
\providecommand \bibnamefont  [1]{#1}%
\providecommand \bibfnamefont [1]{#1}%
\providecommand \citenamefont [1]{#1}%
\providecommand \href@noop [0]{\@secondoftwo}%
\providecommand \href [0]{\begingroup \@sanitize@url \@href}%
\providecommand \@href[1]{\@@startlink{#1}\@@href}%
\providecommand \@@href[1]{\endgroup#1\@@endlink}%
\providecommand \@sanitize@url [0]{\catcode `\\12\catcode `\$12\catcode
  `\&12\catcode `\#12\catcode `\^12\catcode `\_12\catcode `\%12\relax}%
\providecommand \@@startlink[1]{}%
\providecommand \@@endlink[0]{}%
\providecommand \url  [0]{\begingroup\@sanitize@url \@url }%
\providecommand \@url [1]{\endgroup\@href {#1}{\urlprefix }}%
\providecommand \urlprefix  [0]{URL }%
\providecommand \Eprint [0]{\href }%
\providecommand \doibase [0]{http://dx.doi.org/}%
\providecommand \selectlanguage [0]{\@gobble}%
\providecommand \bibinfo  [0]{\@secondoftwo}%
\providecommand \bibfield  [0]{\@secondoftwo}%
\providecommand \translation [1]{[#1]}%
\providecommand \BibitemOpen [0]{}%
\providecommand \bibitemStop [0]{}%
\providecommand \bibitemNoStop [0]{.\EOS\space}%
\providecommand \EOS [0]{\spacefactor3000\relax}%
\providecommand \BibitemShut  [1]{\csname bibitem#1\endcsname}%
\let\auto@bib@innerbib\@empty
\bibitem [{\citenamefont {Tabuchi}\ \emph {et~al.}(2014)\citenamefont
  {Tabuchi}, \citenamefont {Ishino}, \citenamefont {Ishikawa}, \citenamefont
  {Yamazaki}, \citenamefont {Usami},\ and\ \citenamefont
  {Nakamura}}]{Tabuchi_2014}%
  \BibitemOpen
  \bibfield  {author} {\bibinfo {author} {\bibfnamefont {Y.}~\bibnamefont
  {Tabuchi}}, \bibinfo {author} {\bibfnamefont {S.}~\bibnamefont {Ishino}},
  \bibinfo {author} {\bibfnamefont {T.}~\bibnamefont {Ishikawa}}, \bibinfo
  {author} {\bibfnamefont {R.}~\bibnamefont {Yamazaki}}, \bibinfo {author}
  {\bibfnamefont {K.}~\bibnamefont {Usami}}, \ and\ \bibinfo {author}
  {\bibfnamefont {Y.}~\bibnamefont {Nakamura}},\ }\href {\doibase
  10.1103/PhysRevLett.113.083603} {\bibfield  {journal} {\bibinfo  {journal}
  {Phys. Rev. Lett.}\ }\textbf {\bibinfo {volume} {113}},\ \bibinfo {pages}
  {083603} (\bibinfo {year} {2014})}\BibitemShut {NoStop}%
\bibitem [{\citenamefont {Zhang}\ \emph {et~al.}(2014)\citenamefont {Zhang},
  \citenamefont {Zou}, \citenamefont {Jiang},\ and\ \citenamefont
  {Tang}}]{Zhang_2014}%
  \BibitemOpen
  \bibfield  {author} {\bibinfo {author} {\bibfnamefont {X.}~\bibnamefont
  {Zhang}}, \bibinfo {author} {\bibfnamefont {C.-L.}\ \bibnamefont {Zou}},
  \bibinfo {author} {\bibfnamefont {L.}~\bibnamefont {Jiang}}, \ and\ \bibinfo
  {author} {\bibfnamefont {H.~X.}\ \bibnamefont {Tang}},\ }\href {\doibase
  10.1103/PhysRevLett.113.156401} {\bibfield  {journal} {\bibinfo  {journal}
  {Phys. Rev. Lett.}\ }\textbf {\bibinfo {volume} {113}},\ \bibinfo {pages}
  {156401} (\bibinfo {year} {2014})}\BibitemShut {NoStop}%
\bibitem [{\citenamefont {Stancil}\ and\ \citenamefont
  {Prabhakar}(2009)}]{Stancil_Spin_2000}%
  \BibitemOpen
  \bibfield  {author} {\bibinfo {author} {\bibfnamefont {D.}~\bibnamefont
  {Stancil}}\ and\ \bibinfo {author} {\bibfnamefont {A.}~\bibnamefont
  {Prabhakar}},\ }\href@noop {} {}\ (\bibinfo  {publisher} {Springer},\
  \bibinfo {address} {New York},\ \bibinfo {year} {2009})\BibitemShut {NoStop}%
\bibitem [{Fer()}]{FerriSphere}%
  \BibitemOpen
  \href@noop {} {}\bibinfo {howpublished} {\url{http://www.ferrisphere.com/}},\
  \bibinfo {note} {accessed: 2020-05-05}\BibitemShut {NoStop}%
\bibitem [{\citenamefont {Morris}\ \emph {et~al.}(2017)\citenamefont {Morris},
  \citenamefont {van Loo}, \citenamefont {Kosen},\ and\ \citenamefont
  {Karenowska}}]{Morris_2017}%
  \BibitemOpen
  \bibfield  {author} {\bibinfo {author} {\bibfnamefont {R.~G.~E.}\
  \bibnamefont {Morris}}, \bibinfo {author} {\bibfnamefont {A.~F.}\
  \bibnamefont {van Loo}}, \bibinfo {author} {\bibfnamefont {S.}~\bibnamefont
  {Kosen}}, \ and\ \bibinfo {author} {\bibfnamefont {A.~D.}\ \bibnamefont
  {Karenowska}},\ }\href@noop {} {\bibfield  {journal} {\bibinfo  {journal}
  {Scientific Reports}\ }\textbf {\bibinfo {volume} {7}},\ \bibinfo {pages}
  {11511} (\bibinfo {year} {2017})}\BibitemShut {NoStop}%
\bibitem [{\citenamefont {Bai}\ \emph {et~al.}(2015)\citenamefont {Bai},
  \citenamefont {Harder}, \citenamefont {Chen}, \citenamefont {Fan},
  \citenamefont {Xiao},\ and\ \citenamefont {Hu}}]{Bai_Spin_2015}%
  \BibitemOpen
  \bibfield  {author} {\bibinfo {author} {\bibfnamefont {L.}~\bibnamefont
  {Bai}}, \bibinfo {author} {\bibfnamefont {M.}~\bibnamefont {Harder}},
  \bibinfo {author} {\bibfnamefont {Y.~P.}\ \bibnamefont {Chen}}, \bibinfo
  {author} {\bibfnamefont {X.}~\bibnamefont {Fan}}, \bibinfo {author}
  {\bibfnamefont {J.~Q.}\ \bibnamefont {Xiao}}, \ and\ \bibinfo {author}
  {\bibfnamefont {C.-M.}\ \bibnamefont {Hu}},\ }\href@noop {} {\bibfield
  {journal} {\bibinfo  {journal} {Phys. Rev. Lett.}\ }\textbf {\bibinfo
  {volume} {114}},\ \bibinfo {pages} {227201} (\bibinfo {year}
  {2015})}\BibitemShut {NoStop}%
\bibitem [{\citenamefont {Viennot}\ \emph {et~al.}(2015)\citenamefont
  {Viennot}, \citenamefont {Dartiailh}, \citenamefont {Cottet},\ and\
  \citenamefont {Kontos}}]{viennot_coherent_coupling_2015}%
  \BibitemOpen
  \bibfield  {author} {\bibinfo {author} {\bibfnamefont {J.~J.}\ \bibnamefont
  {Viennot}}, \bibinfo {author} {\bibfnamefont {M.~C.}\ \bibnamefont
  {Dartiailh}}, \bibinfo {author} {\bibfnamefont {A.}~\bibnamefont {Cottet}}, \
  and\ \bibinfo {author} {\bibfnamefont {T.}~\bibnamefont {Kontos}},\
  }\href@noop {} {\bibfield  {journal} {\bibinfo  {journal} {Science}\ }\textbf
  {\bibinfo {volume} {349}},\ \bibinfo {pages} {408} (\bibinfo {year}
  {2015})}\BibitemShut {NoStop}%
\bibitem [{\citenamefont {Cao}\ \emph {et~al.}(2015)\citenamefont {Cao},
  \citenamefont {Yan}, \citenamefont {Huebl}, \citenamefont {Goennenwein},\
  and\ \citenamefont {Bauer}}]{Cao_Exchange_2015}%
  \BibitemOpen
  \bibfield  {author} {\bibinfo {author} {\bibfnamefont {Y.}~\bibnamefont
  {Cao}}, \bibinfo {author} {\bibfnamefont {P.}~\bibnamefont {Yan}}, \bibinfo
  {author} {\bibfnamefont {H.}~\bibnamefont {Huebl}}, \bibinfo {author}
  {\bibfnamefont {S.~T.~B.}\ \bibnamefont {Goennenwein}}, \ and\ \bibinfo
  {author} {\bibfnamefont {G.~E.~W.}\ \bibnamefont {Bauer}},\ }\href {\doibase
  10.1103/PhysRevB.91.094423} {\bibfield  {journal} {\bibinfo  {journal} {Phys.
  Rev. B}\ }\textbf {\bibinfo {volume} {91}},\ \bibinfo {pages} {094423}
  (\bibinfo {year} {2015})}\BibitemShut {NoStop}%
\bibitem [{\citenamefont {Zhang}\ \emph {et~al.}(2015)\citenamefont {Zhang},
  \citenamefont {Zou}, \citenamefont {Zhu}, \citenamefont {Marquardt},
  \citenamefont {Jiang},\ and\ \citenamefont {Tang}}]{zhang_magnon_dark_2015}%
  \BibitemOpen
  \bibfield  {author} {\bibinfo {author} {\bibfnamefont {X.}~\bibnamefont
  {Zhang}}, \bibinfo {author} {\bibfnamefont {C.-L.}\ \bibnamefont {Zou}},
  \bibinfo {author} {\bibfnamefont {N.}~\bibnamefont {Zhu}}, \bibinfo {author}
  {\bibfnamefont {F.}~\bibnamefont {Marquardt}}, \bibinfo {author}
  {\bibfnamefont {L.}~\bibnamefont {Jiang}}, \ and\ \bibinfo {author}
  {\bibfnamefont {H.~X.}\ \bibnamefont {Tang}},\ }\href@noop {} {\bibfield
  {journal} {\bibinfo  {journal} {Nature Communications}\ }\textbf {\bibinfo
  {volume} {6}},\ \bibinfo {pages} {8914} (\bibinfo {year} {2015})}\BibitemShut
  {NoStop}%
\bibitem [{\citenamefont {Gloppe}\ \emph {et~al.}(2019)\citenamefont {Gloppe},
  \citenamefont {Hisatomi}, \citenamefont {Nakata}, \citenamefont {Nakamura},\
  and\ \citenamefont {Usami}}]{Gloppe_Resonant_2019}%
  \BibitemOpen
  \bibfield  {author} {\bibinfo {author} {\bibfnamefont {A.}~\bibnamefont
  {Gloppe}}, \bibinfo {author} {\bibfnamefont {R.}~\bibnamefont {Hisatomi}},
  \bibinfo {author} {\bibfnamefont {Y.}~\bibnamefont {Nakata}}, \bibinfo
  {author} {\bibfnamefont {Y.}~\bibnamefont {Nakamura}}, \ and\ \bibinfo
  {author} {\bibfnamefont {K.}~\bibnamefont {Usami}},\ }\href {\doibase
  10.1103/PhysRevApplied.12.014061} {\bibfield  {journal} {\bibinfo  {journal}
  {Phys. Rev. Applied}\ }\textbf {\bibinfo {volume} {12}},\ \bibinfo {pages}
  {014061} (\bibinfo {year} {2019})}\BibitemShut {NoStop}%
\bibitem [{\citenamefont {Goryachev}\ \emph {et~al.}(2018)\citenamefont
  {Goryachev}, \citenamefont {Watt}, \citenamefont {Bourhill}, \citenamefont
  {Kostylev},\ and\ \citenamefont {Tobar}}]{Goryachev_2018}%
  \BibitemOpen
  \bibfield  {author} {\bibinfo {author} {\bibfnamefont {M.}~\bibnamefont
  {Goryachev}}, \bibinfo {author} {\bibfnamefont {S.}~\bibnamefont {Watt}},
  \bibinfo {author} {\bibfnamefont {J.}~\bibnamefont {Bourhill}}, \bibinfo
  {author} {\bibfnamefont {M.}~\bibnamefont {Kostylev}}, \ and\ \bibinfo
  {author} {\bibfnamefont {M.~E.}\ \bibnamefont {Tobar}},\ }\href {\doibase
  10.1103/PhysRevB.97.155129} {\bibfield  {journal} {\bibinfo  {journal} {Phys.
  Rev. B}\ }\textbf {\bibinfo {volume} {97}},\ \bibinfo {pages} {155129}
  (\bibinfo {year} {2018})}\BibitemShut {NoStop}%
\bibitem [{\citenamefont {Bhoi}\ \emph {et~al.}(2014)\citenamefont {Bhoi},
  \citenamefont {Cliff}, \citenamefont {Maksymov}, \citenamefont {Kostylev},
  \citenamefont {Aiyar}, \citenamefont {Venkataramani}, \citenamefont
  {Prasad},\ and\ \citenamefont {Stamps}}]{Bhoi_2014}%
  \BibitemOpen
  \bibfield  {author} {\bibinfo {author} {\bibfnamefont {B.}~\bibnamefont
  {Bhoi}}, \bibinfo {author} {\bibfnamefont {T.}~\bibnamefont {Cliff}},
  \bibinfo {author} {\bibfnamefont {I.~S.}\ \bibnamefont {Maksymov}}, \bibinfo
  {author} {\bibfnamefont {M.}~\bibnamefont {Kostylev}}, \bibinfo {author}
  {\bibfnamefont {R.}~\bibnamefont {Aiyar}}, \bibinfo {author} {\bibfnamefont
  {N.}~\bibnamefont {Venkataramani}}, \bibinfo {author} {\bibfnamefont
  {S.}~\bibnamefont {Prasad}}, \ and\ \bibinfo {author} {\bibfnamefont {R.~L.}\
  \bibnamefont {Stamps}},\ }\href {\doibase 10.1063/1.4904857} {\bibfield
  {journal} {\bibinfo  {journal} {Journal of Applied Physics}\ }\textbf
  {\bibinfo {volume} {116}},\ \bibinfo {pages} {243906} (\bibinfo {year}
  {2014})}\BibitemShut {NoStop}%
\bibitem [{\citenamefont {Li}\ \emph {et~al.}(2019)\citenamefont {Li},
  \citenamefont {Polakovic}, \citenamefont {Wang}, \citenamefont {Xu},
  \citenamefont {Lendinez}, \citenamefont {Zhang}, \citenamefont {Ding},
  \citenamefont {Khaire}, \citenamefont {Saglam}, \citenamefont {Divan},
  \citenamefont {Pearson}, \citenamefont {Kwok}, \citenamefont {Xiao},
  \citenamefont {Novosad}, \citenamefont {Hoffmann},\ and\ \citenamefont
  {Zhang}}]{Li_2019}%
  \BibitemOpen
  \bibfield  {author} {\bibinfo {author} {\bibfnamefont {Y.}~\bibnamefont
  {Li}}, \bibinfo {author} {\bibfnamefont {T.}~\bibnamefont {Polakovic}},
  \bibinfo {author} {\bibfnamefont {Y.-L.}\ \bibnamefont {Wang}}, \bibinfo
  {author} {\bibfnamefont {J.}~\bibnamefont {Xu}}, \bibinfo {author}
  {\bibfnamefont {S.}~\bibnamefont {Lendinez}}, \bibinfo {author}
  {\bibfnamefont {Z.}~\bibnamefont {Zhang}}, \bibinfo {author} {\bibfnamefont
  {J.}~\bibnamefont {Ding}}, \bibinfo {author} {\bibfnamefont {T.}~\bibnamefont
  {Khaire}}, \bibinfo {author} {\bibfnamefont {H.}~\bibnamefont {Saglam}},
  \bibinfo {author} {\bibfnamefont {R.}~\bibnamefont {Divan}}, \bibinfo
  {author} {\bibfnamefont {J.}~\bibnamefont {Pearson}}, \bibinfo {author}
  {\bibfnamefont {W.-K.}\ \bibnamefont {Kwok}}, \bibinfo {author}
  {\bibfnamefont {Z.}~\bibnamefont {Xiao}}, \bibinfo {author} {\bibfnamefont
  {V.}~\bibnamefont {Novosad}}, \bibinfo {author} {\bibfnamefont
  {A.}~\bibnamefont {Hoffmann}}, \ and\ \bibinfo {author} {\bibfnamefont
  {W.}~\bibnamefont {Zhang}},\ }\href {\doibase 10.1103/PhysRevLett.123.107701}
  {\bibfield  {journal} {\bibinfo  {journal} {Phys. Rev. Lett.}\ }\textbf
  {\bibinfo {volume} {123}},\ \bibinfo {pages} {107701} (\bibinfo {year}
  {2019})}\BibitemShut {NoStop}%
\bibitem [{\citenamefont {Harder}\ \emph {et~al.}(2018)\citenamefont {Harder},
  \citenamefont {Yang}, \citenamefont {Yao}, \citenamefont {Yu}, \citenamefont
  {Rao}, \citenamefont {Gui}, \citenamefont {Stamps},\ and\ \citenamefont
  {Hu}}]{Harder_2018}%
  \BibitemOpen
  \bibfield  {author} {\bibinfo {author} {\bibfnamefont {M.}~\bibnamefont
  {Harder}}, \bibinfo {author} {\bibfnamefont {Y.}~\bibnamefont {Yang}},
  \bibinfo {author} {\bibfnamefont {B.~M.}\ \bibnamefont {Yao}}, \bibinfo
  {author} {\bibfnamefont {C.~H.}\ \bibnamefont {Yu}}, \bibinfo {author}
  {\bibfnamefont {J.~W.}\ \bibnamefont {Rao}}, \bibinfo {author} {\bibfnamefont
  {Y.~S.}\ \bibnamefont {Gui}}, \bibinfo {author} {\bibfnamefont {R.~L.}\
  \bibnamefont {Stamps}}, \ and\ \bibinfo {author} {\bibfnamefont {C.-M.}\
  \bibnamefont {Hu}},\ }\href {\doibase 10.1103/PhysRevLett.121.137203}
  {\bibfield  {journal} {\bibinfo  {journal} {Phys. Rev. Lett.}\ }\textbf
  {\bibinfo {volume} {121}},\ \bibinfo {pages} {137203} (\bibinfo {year}
  {2018})}\BibitemShut {NoStop}%
\bibitem [{\citenamefont {Wang}\ \emph {et~al.}(2019)\citenamefont {Wang},
  \citenamefont {Rao}, \citenamefont {Yang}, \citenamefont {Xu}, \citenamefont
  {Gui}, \citenamefont {Yao}, \citenamefont {You},\ and\ \citenamefont
  {Hu}}]{Wang_2019}%
  \BibitemOpen
  \bibfield  {author} {\bibinfo {author} {\bibfnamefont {Y.-P.}\ \bibnamefont
  {Wang}}, \bibinfo {author} {\bibfnamefont {J.~W.}\ \bibnamefont {Rao}},
  \bibinfo {author} {\bibfnamefont {Y.}~\bibnamefont {Yang}}, \bibinfo {author}
  {\bibfnamefont {P.-C.}\ \bibnamefont {Xu}}, \bibinfo {author} {\bibfnamefont
  {Y.~S.}\ \bibnamefont {Gui}}, \bibinfo {author} {\bibfnamefont {B.~M.}\
  \bibnamefont {Yao}}, \bibinfo {author} {\bibfnamefont {J.~Q.}\ \bibnamefont
  {You}}, \ and\ \bibinfo {author} {\bibfnamefont {C.-M.}\ \bibnamefont {Hu}},\
  }\href {\doibase 10.1103/PhysRevLett.123.127202} {\bibfield  {journal}
  {\bibinfo  {journal} {Phys. Rev. Lett.}\ }\textbf {\bibinfo {volume} {123}},\
  \bibinfo {pages} {127202} (\bibinfo {year} {2019})}\BibitemShut {NoStop}%
\bibitem [{\citenamefont {Zhang}\ \emph {et~al.}(2016)\citenamefont {Zhang},
  \citenamefont {Zou}, \citenamefont {Jiang},\ and\ \citenamefont
  {Tang}}]{Zhang_2016}%
  \BibitemOpen
  \bibfield  {author} {\bibinfo {author} {\bibfnamefont {X.}~\bibnamefont
  {Zhang}}, \bibinfo {author} {\bibfnamefont {C.-L.}\ \bibnamefont {Zou}},
  \bibinfo {author} {\bibfnamefont {L.}~\bibnamefont {Jiang}}, \ and\ \bibinfo
  {author} {\bibfnamefont {H.~X.}\ \bibnamefont {Tang}},\ }\href@noop {}
  {\bibfield  {journal} {\bibinfo  {journal} {Science Advances}\ }\textbf
  {\bibinfo {volume} {2}} (\bibinfo {year} {2016})}\BibitemShut {NoStop}%
\bibitem [{\citenamefont {Potts}\ \emph {et~al.}(2020)\citenamefont {Potts},
  \citenamefont {Bittencourt}, \citenamefont {Kusminskiy},\ and\ \citenamefont
  {Davis}}]{Potts_2020}%
  \BibitemOpen
  \bibfield  {author} {\bibinfo {author} {\bibfnamefont {C.}~\bibnamefont
  {Potts}}, \bibinfo {author} {\bibfnamefont {V.}~\bibnamefont {Bittencourt}},
  \bibinfo {author} {\bibfnamefont {S.~V.}\ \bibnamefont {Kusminskiy}}, \ and\
  \bibinfo {author} {\bibfnamefont {J.}~\bibnamefont {Davis}},\ }\href
  {\doibase 10.1103/PhysRevApplied.13.064001} {\bibfield  {journal} {\bibinfo
  {journal} {Phys. Rev. Applied}\ }\textbf {\bibinfo {volume} {13}},\ \bibinfo
  {pages} {064001} (\bibinfo {year} {2020})}\BibitemShut {NoStop}%
\bibitem [{\citenamefont {Tabuchi}\ \emph {et~al.}(2015)\citenamefont
  {Tabuchi}, \citenamefont {Ishino}, \citenamefont {Noguchi}, \citenamefont
  {Ishikawa}, \citenamefont {Yamazaki}, \citenamefont {Usami},\ and\
  \citenamefont {Nakamura}}]{Tabuchi_2015}%
  \BibitemOpen
  \bibfield  {author} {\bibinfo {author} {\bibfnamefont {Y.}~\bibnamefont
  {Tabuchi}}, \bibinfo {author} {\bibfnamefont {S.}~\bibnamefont {Ishino}},
  \bibinfo {author} {\bibfnamefont {A.}~\bibnamefont {Noguchi}}, \bibinfo
  {author} {\bibfnamefont {T.}~\bibnamefont {Ishikawa}}, \bibinfo {author}
  {\bibfnamefont {R.}~\bibnamefont {Yamazaki}}, \bibinfo {author}
  {\bibfnamefont {K.}~\bibnamefont {Usami}}, \ and\ \bibinfo {author}
  {\bibfnamefont {Y.}~\bibnamefont {Nakamura}},\ }\href {\doibase
  10.1126/science.aaa3693} {\bibfield  {journal} {\bibinfo  {journal}
  {Science}\ }\textbf {\bibinfo {volume} {349}},\ \bibinfo {pages} {405}
  (\bibinfo {year} {2015})}\BibitemShut {NoStop}%
\bibitem [{\citenamefont {Hisatomi}\ \emph {et~al.}(2016)\citenamefont
  {Hisatomi}, \citenamefont {Osada}, \citenamefont {Tabuchi}, \citenamefont
  {Ishikawa}, \citenamefont {Noguchi}, \citenamefont {Yamazaki}, \citenamefont
  {Usami},\ and\ \citenamefont {Nakamura}}]{Hisatomi_2016}%
  \BibitemOpen
  \bibfield  {author} {\bibinfo {author} {\bibfnamefont {R.}~\bibnamefont
  {Hisatomi}}, \bibinfo {author} {\bibfnamefont {A.}~\bibnamefont {Osada}},
  \bibinfo {author} {\bibfnamefont {Y.}~\bibnamefont {Tabuchi}}, \bibinfo
  {author} {\bibfnamefont {T.}~\bibnamefont {Ishikawa}}, \bibinfo {author}
  {\bibfnamefont {A.}~\bibnamefont {Noguchi}}, \bibinfo {author} {\bibfnamefont
  {R.}~\bibnamefont {Yamazaki}}, \bibinfo {author} {\bibfnamefont
  {K.}~\bibnamefont {Usami}}, \ and\ \bibinfo {author} {\bibfnamefont
  {Y.}~\bibnamefont {Nakamura}},\ }\href {\doibase 10.1103/PhysRevB.93.174427}
  {\bibfield  {journal} {\bibinfo  {journal} {Phys. Rev. B}\ }\textbf {\bibinfo
  {volume} {93}},\ \bibinfo {pages} {174427} (\bibinfo {year}
  {2016})}\BibitemShut {NoStop}%
\bibitem [{\citenamefont {Zhu}\ \emph {et~al.}()\citenamefont {Zhu},
  \citenamefont {Zhang}, \citenamefont {Han}, \citenamefont {Zou},
  \citenamefont {Zhong}, \citenamefont {Wang}, \citenamefont {Jiang},\ and\
  \citenamefont {Tang}}]{Zhu_2020}%
  \BibitemOpen
  \bibfield  {author} {\bibinfo {author} {\bibfnamefont {N.}~\bibnamefont
  {Zhu}}, \bibinfo {author} {\bibfnamefont {X.}~\bibnamefont {Zhang}}, \bibinfo
  {author} {\bibfnamefont {X.}~\bibnamefont {Han}}, \bibinfo {author}
  {\bibfnamefont {C.-L.}\ \bibnamefont {Zou}}, \bibinfo {author} {\bibfnamefont
  {C.}~\bibnamefont {Zhong}}, \bibinfo {author} {\bibfnamefont {C.-H.}\
  \bibnamefont {Wang}}, \bibinfo {author} {\bibfnamefont {L.}~\bibnamefont
  {Jiang}}, \ and\ \bibinfo {author} {\bibfnamefont {H.~X.}\ \bibnamefont
  {Tang}},\ }\href@noop {} {}\Eprint {http://arxiv.org/abs/2005.06429}
  {arXiv:2005.06429} \BibitemShut {NoStop}%
\bibitem [{\citenamefont {Lachance-Quirion}\ \emph {et~al.}(2020)\citenamefont
  {Lachance-Quirion}, \citenamefont {Wolski}, \citenamefont {Tabuchi},
  \citenamefont {Kono}, \citenamefont {Usami},\ and\ \citenamefont
  {Nakamura}}]{Lachance-Quirion_2020}%
  \BibitemOpen
  \bibfield  {author} {\bibinfo {author} {\bibfnamefont {D.}~\bibnamefont
  {Lachance-Quirion}}, \bibinfo {author} {\bibfnamefont {S.~P.}\ \bibnamefont
  {Wolski}}, \bibinfo {author} {\bibfnamefont {Y.}~\bibnamefont {Tabuchi}},
  \bibinfo {author} {\bibfnamefont {S.}~\bibnamefont {Kono}}, \bibinfo {author}
  {\bibfnamefont {K.}~\bibnamefont {Usami}}, \ and\ \bibinfo {author}
  {\bibfnamefont {Y.}~\bibnamefont {Nakamura}},\ }\href {\doibase
  10.1126/science.aaz9236} {\bibfield  {journal} {\bibinfo  {journal}
  {Science}\ }\textbf {\bibinfo {volume} {367}},\ \bibinfo {pages} {425}
  (\bibinfo {year} {2020})}\BibitemShut {NoStop}%
\bibitem [{\citenamefont {Flower}\ \emph
  {et~al.}(2019{\natexlab{a}})\citenamefont {Flower}, \citenamefont {Bourhill},
  \citenamefont {Goryachev},\ and\ \citenamefont {Tobar}}]{Flower_2019}%
  \BibitemOpen
  \bibfield  {author} {\bibinfo {author} {\bibfnamefont {G.}~\bibnamefont
  {Flower}}, \bibinfo {author} {\bibfnamefont {J.}~\bibnamefont {Bourhill}},
  \bibinfo {author} {\bibfnamefont {M.}~\bibnamefont {Goryachev}}, \ and\
  \bibinfo {author} {\bibfnamefont {M.~E.}\ \bibnamefont {Tobar}},\ }\href
  {\doibase https://doi.org/10.1016/j.dark.2019.100306} {\bibfield  {journal}
  {\bibinfo  {journal} {Physics of the Dark Universe}\ }\textbf {\bibinfo
  {volume} {25}},\ \bibinfo {pages} {100306} (\bibinfo {year}
  {2019}{\natexlab{a}})}\BibitemShut {NoStop}%
\bibitem [{\citenamefont {Crescini}\ \emph {et~al.}(2020)\citenamefont
  {Crescini}, \citenamefont {Alesini}, \citenamefont {Braggio}, \citenamefont
  {Carugno}, \citenamefont {D'Agostino}, \citenamefont {Di~Gioacchino},
  \citenamefont {Falferi}, \citenamefont {Gambardella}, \citenamefont {Gatti},
  \citenamefont {Iannone}, \citenamefont {Ligi}, \citenamefont {Lombardi},
  \citenamefont {Ortolan}, \citenamefont {Pengo}, \citenamefont {Ruoso},\ and\
  \citenamefont {Taffarello}}]{Crescini_2020}%
  \BibitemOpen
  \bibfield  {author} {\bibinfo {author} {\bibfnamefont {N.}~\bibnamefont
  {Crescini}}, \bibinfo {author} {\bibfnamefont {D.}~\bibnamefont {Alesini}},
  \bibinfo {author} {\bibfnamefont {C.}~\bibnamefont {Braggio}}, \bibinfo
  {author} {\bibfnamefont {G.}~\bibnamefont {Carugno}}, \bibinfo {author}
  {\bibfnamefont {D.}~\bibnamefont {D'Agostino}}, \bibinfo {author}
  {\bibfnamefont {D.}~\bibnamefont {Di~Gioacchino}}, \bibinfo {author}
  {\bibfnamefont {P.}~\bibnamefont {Falferi}}, \bibinfo {author} {\bibfnamefont
  {U.}~\bibnamefont {Gambardella}}, \bibinfo {author} {\bibfnamefont
  {C.}~\bibnamefont {Gatti}}, \bibinfo {author} {\bibfnamefont
  {G.}~\bibnamefont {Iannone}}, \bibinfo {author} {\bibfnamefont
  {C.}~\bibnamefont {Ligi}}, \bibinfo {author} {\bibfnamefont {A.}~\bibnamefont
  {Lombardi}}, \bibinfo {author} {\bibfnamefont {A.}~\bibnamefont {Ortolan}},
  \bibinfo {author} {\bibfnamefont {R.}~\bibnamefont {Pengo}}, \bibinfo
  {author} {\bibfnamefont {G.}~\bibnamefont {Ruoso}}, \ and\ \bibinfo {author}
  {\bibfnamefont {L.}~\bibnamefont {Taffarello}} (\bibinfo {collaboration}
  {QUAX Collaboration}),\ }\href {\doibase 10.1103/PhysRevLett.124.171801}
  {\bibfield  {journal} {\bibinfo  {journal} {Phys. Rev. Lett.}\ }\textbf
  {\bibinfo {volume} {124}},\ \bibinfo {pages} {171801} (\bibinfo {year}
  {2020})}\BibitemShut {NoStop}%
\bibitem [{\citenamefont {Hyde}\ \emph {et~al.}(2017)\citenamefont {Hyde},
  \citenamefont {Bai}, \citenamefont {Harder}, \citenamefont {Dyck},\ and\
  \citenamefont {Hu}}]{Hyde_2017}%
  \BibitemOpen
  \bibfield  {author} {\bibinfo {author} {\bibfnamefont {P.}~\bibnamefont
  {Hyde}}, \bibinfo {author} {\bibfnamefont {L.}~\bibnamefont {Bai}}, \bibinfo
  {author} {\bibfnamefont {M.}~\bibnamefont {Harder}}, \bibinfo {author}
  {\bibfnamefont {C.}~\bibnamefont {Dyck}}, \ and\ \bibinfo {author}
  {\bibfnamefont {C.-M.}\ \bibnamefont {Hu}},\ }\href {\doibase
  10.1103/PhysRevB.95.094416} {\bibfield  {journal} {\bibinfo  {journal} {Phys.
  Rev. B}\ }\textbf {\bibinfo {volume} {95}},\ \bibinfo {pages} {094416}
  (\bibinfo {year} {2017})}\BibitemShut {NoStop}%
\bibitem [{\citenamefont {Bourhill}\ \emph {et~al.}()\citenamefont {Bourhill},
  \citenamefont {Castel}, \citenamefont {Manchec},\ and\ \citenamefont
  {Cochet}}]{Bourhill_2019_spectroscopy}%
  \BibitemOpen
  \bibfield  {author} {\bibinfo {author} {\bibfnamefont {J.}~\bibnamefont
  {Bourhill}}, \bibinfo {author} {\bibfnamefont {V.}~\bibnamefont {Castel}},
  \bibinfo {author} {\bibfnamefont {A.}~\bibnamefont {Manchec}}, \ and\
  \bibinfo {author} {\bibfnamefont {G.}~\bibnamefont {Cochet}},\ }\href@noop {}
  {}\Eprint {http://arxiv.org/abs/1910.08333} {arXiv:1910.08333} \BibitemShut
  {NoStop}%
\bibitem [{\citenamefont {Clark}\ \emph {et~al.}(2018)\citenamefont {Clark},
  \citenamefont {Vadakkumbatt}, \citenamefont {Souris}, \citenamefont {Ramp},\
  and\ \citenamefont {Davis}}]{Clark_2018}%
  \BibitemOpen
  \bibfield  {author} {\bibinfo {author} {\bibfnamefont {T.~J.}\ \bibnamefont
  {Clark}}, \bibinfo {author} {\bibfnamefont {V.}~\bibnamefont {Vadakkumbatt}},
  \bibinfo {author} {\bibfnamefont {F.}~\bibnamefont {Souris}}, \bibinfo
  {author} {\bibfnamefont {H.}~\bibnamefont {Ramp}}, \ and\ \bibinfo {author}
  {\bibfnamefont {J.~P.}\ \bibnamefont {Davis}},\ }\href {\doibase
  10.1063/1.5051042} {\bibfield  {journal} {\bibinfo  {journal} {Review of
  Scientific Instruments}\ }\textbf {\bibinfo {volume} {89}},\ \bibinfo {pages}
  {114704} (\bibinfo {year} {2018})}\BibitemShut {NoStop}%
\bibitem [{\citenamefont {Brecht}\ \emph {et~al.}(2015)\citenamefont {Brecht},
  \citenamefont {Reagor}, \citenamefont {Chu}, \citenamefont {Pfaff},
  \citenamefont {Wang}, \citenamefont {Frunzio}, \citenamefont {Devoret},\ and\
  \citenamefont {Schoelkopf}}]{Brecht_2015}%
  \BibitemOpen
  \bibfield  {author} {\bibinfo {author} {\bibfnamefont {T.}~\bibnamefont
  {Brecht}}, \bibinfo {author} {\bibfnamefont {M.}~\bibnamefont {Reagor}},
  \bibinfo {author} {\bibfnamefont {Y.}~\bibnamefont {Chu}}, \bibinfo {author}
  {\bibfnamefont {W.}~\bibnamefont {Pfaff}}, \bibinfo {author} {\bibfnamefont
  {C.}~\bibnamefont {Wang}}, \bibinfo {author} {\bibfnamefont {L.}~\bibnamefont
  {Frunzio}}, \bibinfo {author} {\bibfnamefont {M.~H.}\ \bibnamefont
  {Devoret}}, \ and\ \bibinfo {author} {\bibfnamefont {R.~J.}\ \bibnamefont
  {Schoelkopf}},\ }\href {\doibase 10.1063/1.4935541} {\bibfield  {journal}
  {\bibinfo  {journal} {Applied Physics Letters}\ }\textbf {\bibinfo {volume}
  {107}},\ \bibinfo {pages} {192603} (\bibinfo {year} {2015})}\BibitemShut
  {NoStop}%
\bibitem [{\citenamefont {Walker}(1958)}]{Walker_1958}%
  \BibitemOpen
  \bibfield  {author} {\bibinfo {author} {\bibfnamefont {L.~R.}\ \bibnamefont
  {Walker}},\ }\href {\doibase 10.1063/1.1723117} {\bibfield  {journal}
  {\bibinfo  {journal} {Journal of Applied Physics}\ }\textbf {\bibinfo
  {volume} {29}},\ \bibinfo {pages} {318} (\bibinfo {year} {1958})}\BibitemShut
  {NoStop}%
\bibitem [{\citenamefont {Fletcher}\ and\ \citenamefont
  {Bell}(1959)}]{Fletcher_1959}%
  \BibitemOpen
  \bibfield  {author} {\bibinfo {author} {\bibfnamefont {P.~C.}\ \bibnamefont
  {Fletcher}}\ and\ \bibinfo {author} {\bibfnamefont {R.~O.}\ \bibnamefont
  {Bell}},\ }\href {\doibase 10.1063/1.1735216} {\bibfield  {journal} {\bibinfo
   {journal} {Journal of Applied Physics}\ }\textbf {\bibinfo {volume} {30}},\
  \bibinfo {pages} {687} (\bibinfo {year} {1959})}\BibitemShut {NoStop}%
\bibitem [{\citenamefont {Holstein}\ and\ \citenamefont
  {Primakoff}(1940)}]{Holstein_1940}%
  \BibitemOpen
  \bibfield  {author} {\bibinfo {author} {\bibfnamefont {T.}~\bibnamefont
  {Holstein}}\ and\ \bibinfo {author} {\bibfnamefont {H.}~\bibnamefont
  {Primakoff}},\ }\href {\doibase 10.1103/PhysRev.58.1098} {\bibfield
  {journal} {\bibinfo  {journal} {Phys. Rev.}\ }\textbf {\bibinfo {volume}
  {58}},\ \bibinfo {pages} {1098} (\bibinfo {year} {1940})}\BibitemShut
  {NoStop}%
\bibitem [{\citenamefont {Flower}\ \emph
  {et~al.}(2019{\natexlab{b}})\citenamefont {Flower}, \citenamefont
  {Goryachev}, \citenamefont {Bourhill},\ and\ \citenamefont
  {Tobar}}]{Flower_Experimental_2019}%
  \BibitemOpen
  \bibfield  {author} {\bibinfo {author} {\bibfnamefont {G.}~\bibnamefont
  {Flower}}, \bibinfo {author} {\bibfnamefont {M.}~\bibnamefont {Goryachev}},
  \bibinfo {author} {\bibfnamefont {J.}~\bibnamefont {Bourhill}}, \ and\
  \bibinfo {author} {\bibfnamefont {M.~E.}\ \bibnamefont {Tobar}},\ }\href
  {https://iopscience.iop.org/article/10.1088/1367-2630/ab3e1c5} {\bibfield
  {journal} {\bibinfo  {journal} {New J. Phys.}\ }\textbf {\bibinfo {volume}
  {21}},\ \bibinfo {pages} {095004} (\bibinfo {year}
  {2019}{\natexlab{b}})}\BibitemShut {NoStop}%
\bibitem [{\citenamefont {Gardiner}\ and\ \citenamefont
  {Zoller}(2000)}]{gardiner_quantum_2000}%
  \BibitemOpen
  \bibfield  {author} {\bibinfo {author} {\bibfnamefont {C.~W.}\ \bibnamefont
  {Gardiner}}\ and\ \bibinfo {author} {\bibfnamefont {P.}~\bibnamefont
  {Zoller}},\ }\href@noop {} {}\bibinfo {edition} {2nd}\ ed.\ (\bibinfo
  {publisher} {{Springer}},\ \bibinfo {address} {{Berlin}},\ \bibinfo {year}
  {2000})\BibitemShut {NoStop}%
\bibitem [{\citenamefont {Zare~Rameshti}, \citenamefont {Cao},\ and\
  \citenamefont {Bauer}(2015)}]{Rameshti_2015}%
  \BibitemOpen
  \bibfield  {author} {\bibinfo {author} {\bibfnamefont {B.}~\bibnamefont
  {Zare~Rameshti}}, \bibinfo {author} {\bibfnamefont {Y.}~\bibnamefont {Cao}},
  \ and\ \bibinfo {author} {\bibfnamefont {G.~E.~W.}\ \bibnamefont {Bauer}},\
  }\href {\doibase 10.1103/PhysRevB.91.214430} {\bibfield  {journal} {\bibinfo
  {journal} {Phys. Rev. B}\ }\textbf {\bibinfo {volume} {91}},\ \bibinfo
  {pages} {214430} (\bibinfo {year} {2015})}\BibitemShut {NoStop}%
\bibitem [{\citenamefont {Bourhill}\ \emph {et~al.}(2016)\citenamefont
  {Bourhill}, \citenamefont {Kostylev}, \citenamefont {Goryachev},
  \citenamefont {Creedon},\ and\ \citenamefont {Tobar}}]{Bourhill_2016}%
  \BibitemOpen
  \bibfield  {author} {\bibinfo {author} {\bibfnamefont {J.}~\bibnamefont
  {Bourhill}}, \bibinfo {author} {\bibfnamefont {N.}~\bibnamefont {Kostylev}},
  \bibinfo {author} {\bibfnamefont {M.}~\bibnamefont {Goryachev}}, \bibinfo
  {author} {\bibfnamefont {D.~L.}\ \bibnamefont {Creedon}}, \ and\ \bibinfo
  {author} {\bibfnamefont {M.~E.}\ \bibnamefont {Tobar}},\ }\href {\doibase
  10.1103/PhysRevB.93.144420} {\bibfield  {journal} {\bibinfo  {journal} {Phys.
  Rev. B}\ }\textbf {\bibinfo {volume} {93}},\ \bibinfo {pages} {144420}
  (\bibinfo {year} {2016})}\BibitemShut {NoStop}%
\bibitem [{\citenamefont {Goryachev}\ \emph {et~al.}(2014)\citenamefont
  {Goryachev}, \citenamefont {Farr}, \citenamefont {Creedon}, \citenamefont
  {Fan}, \citenamefont {Kostylev},\ and\ \citenamefont
  {Tobar}}]{Goryachev_2014}%
  \BibitemOpen
  \bibfield  {author} {\bibinfo {author} {\bibfnamefont {M.}~\bibnamefont
  {Goryachev}}, \bibinfo {author} {\bibfnamefont {W.~G.}\ \bibnamefont {Farr}},
  \bibinfo {author} {\bibfnamefont {D.~L.}\ \bibnamefont {Creedon}}, \bibinfo
  {author} {\bibfnamefont {Y.}~\bibnamefont {Fan}}, \bibinfo {author}
  {\bibfnamefont {M.}~\bibnamefont {Kostylev}}, \ and\ \bibinfo {author}
  {\bibfnamefont {M.~E.}\ \bibnamefont {Tobar}},\ }\href {\doibase
  10.1103/PhysRevApplied.2.054002} {\bibfield  {journal} {\bibinfo  {journal}
  {Phys. Rev. Applied}\ }\textbf {\bibinfo {volume} {2}},\ \bibinfo {pages}
  {054002} (\bibinfo {year} {2014})}\BibitemShut {NoStop}%
\end{thebibliography}%

\end{document}